\documentclass[%
 preprint,
 amsmath,amssymb,
 aps,
onecolumn
]{revtex4-1}

\usepackage{graphicx}
\usepackage{dcolumn}
\usepackage{bm}

\begin{document}

\title{A model to inter-relate the values of the quantum electrodynamic, gravitational and cosmological constants
}

\author{L. Eaves}
\affiliation{School of Physics \& Astronomy, University of Nottingham, Nottingham NG7 2RD, UK  \\
School of Physics \& Astronomy, University of Manchester, Manchester M13 9PL, UK
}

\date{\today}

\begin{abstract}
The fundamental constants of electromagnetism, gravity and quantum mechanics can be related empirically by the numerical approximation $\ln(V_e/V_P)\approx \alpha^{-1}$, where $\alpha$ is the low energy value of the electromagnetic fine structure constant and $V_e$ and $V_P$ are volumes corresponding to the classical electron radius, $r_e$, and the Planck length respectively.  This logarithmic relation is used in an ideal gas model to determine the work, $W$, done when a hypothetical vacuum fluctuation expands relativistically from $V_P$ to $V_e$ in a time limited by the uncertainty principle.  It is proposed that the expansion is a phenomenological representation of a quantum transition from a Planck-scale initial state into a final virtual photonic state of energy $W\simeq \hbar c/2r_e$ and lifetime $\simeq r_e/c$, occupying a volume $\simeq V_e$.  The magnitude of the negative gravitational self-energy density, $\rho_G$, of this virtual state is found to be within $\sim 10\%$ of the measured value of the positive ``dark energy'' density, $\rho_\Lambda$.  It is proposed that this is not merely an ``accidental'' numerical coincidence but has physical significance, namely that the sum of the two energy densities is zero, i.e. $\rho_\Lambda+\rho_G=0$.  This relation gives a value of the cosmological constant, $\Lambda$, in agreement with astronomical measurements.   The implications of these inter-relations between $\Lambda$, the gravitational constant, $G$, and $\alpha$ are outlined.

\end{abstract}

\maketitle


\section{Introduction}

The unification of gravity with the electroweak and strong interactions remains a major goal in physics. Of particular interest is the relation between the electromagnetic fine structure constant and Newton's gravitational constant, $G$ \cite{1DeWitt,2Isham,3Carr,4Rozental,5Ellis,6Barrow,7Page,8Jentschura,9Eaves}.  A second unresolved problem is the very small but finite and positive value of the cosmological constant $\Lambda$ \cite{10Riess,11Perlmutter,12Riess,13Weinberg,14Weinberg,15Efstathiou,16Efstathiou,17Beck,18Barrow,19Shaw}.  Here a model is proposed to address these problems.

A logarithmic relation of the form, $\alpha^{-1}\sim\ln\alpha_G^{-1}$, has long been regarded as a requirement for self-consistent electrodynamics \cite{1DeWitt,2Isham,3Carr,4Rozental,5Ellis,6Barrow,7Page,8Jentschura}; here $\alpha$ is the value of the electromagnetic fine structure constant in the low energy limit, $\alpha_G=Gm_N^2/\hbar c$ is the conventionally defined form of the gravitational coupling constant and $m_N$ is the proton mass.  The relation between cosmological parameters and the large exponential number $e^{1/\alpha}\simeq 3.27\times 10^{59}$ has been discussed by Barrow and Tipler \cite{6Barrow}, and a renormalisation group analysis of supersymmetric grand unified theories has been used by  Page \cite{7Page} to deduce that $\alpha^{-1}\approx(5/\pi)\ln\alpha_G^{-1}$.  The following empirical, yet numerically accurate, approximation has also been suggested recently \cite{9Eaves} as an alternative way of relating $G$ to $\alpha$ in terms of two fundamental length scales: 
\begin{equation}
\left(\frac{r_e}{L_P}\right)^2=\frac{\alpha q^2}{2\pi G m_e^2}\approx e^{2/3\alpha}.  
\label{eq:1}
\end{equation}

Here, $r_e=e^2/4\pi\varepsilon_0 m_e c^2=2.82\times10^{-15}$ m is the classical electron radius, $L_p=\sqrt{h G/c^3}=4.05\times 10^{-35}$ m is the Planck length, $e$ is the quantum of electrical charge, $m_e$ is the electron rest mass,  $\alpha=e^2/ 4\pi\epsilon_0\hbar c$ and $q^2=e^2/ 4\pi\varepsilon_0$.  Relation (\ref{eq:1}) involves $m_e$ and so avoids using the mass of the nucleon, a particle with a complex internal structure.  With the other constants set to their experimentally measured values, relation (\ref{eq:1}) would become an equality if the numerical value of $\alpha^{-1}$ were to be $137.066$.  This number exceeds by only $\sim 0.02\%$ the actual measured value of $\alpha^{-1}$ ($\alpha^{-1}=137.036$, rounded to 6 significant figures). It is proposed that relation (\ref{eq:1}) is not merely a numerical coincidence but has a physical significance that can be used to inter-relate accurately the values of $\alpha$, $G$ and $\Lambda$, as described in the next section. 

\section{A model to interrelate $\alpha$, $G$ and $\Lambda$}

It was noted in reference \cite{9Eaves} that relation (\ref{eq:1}) can be expressed in integral form, in terms of the classical electron volume, $V_e$, and the Planck volume, $V_P$:

\begin{equation}
\ln \left( \frac{V_e}{V_P}\right) = \int^{V_e}_{V_P}V^{-1}dV\approx \alpha^{-1}.
\label{eq:VeVp}
\end{equation}

Such a volume integral is familiar and suggestive as it appears in the calculation of the work done, $W$, by a classical ideal gas when it expands isothermally.  With this in mind, a phenomenological model is now proposed to explain the physical significance of relations (\ref{eq:1}) and (\ref{eq:VeVp}) and to derive an expression for $\Lambda$ in terms of $G$ and the quantum electrodynamic constants, $\alpha$, $m_e$ and $e$.

In the absence of a quantum theory of gravity and electrodynamics at the Plank scale, relation (\ref{eq:VeVp}) is combined with the uncertainty principle to model the transition of a vacuum fluctuation between the Plank and electromagnetic length scales.  This is achieved by representing it in terms of the expansion of an isotropic fluid with a simple ideal gas equation of state given by
\begin{equation}
PV= jm_e c^2.
\label{eq:PV}
\end{equation}

Here $P$ and $V$ are the fluid's pressure and volume respectively.  Since relations (\ref{eq:1}) and (\ref{eq:VeVp}) involve $m_e$ and $V_e$, the constant term on the right-hand side of relation (\ref{eq:PV}) is expressed in units of the electron's rest mass energy as $jm_ec^2$, where $j$ is a dimensionless parameter whose numerical value will be considered next.  In effect, this energy takes the place of the thermal energy of a classical ideal gas obeying Boyle's law.  Using relation (\ref{eq:VeVp}), the work, $W$, done by the fluid during an expansion from $V_P$ to $V_e$ is given by 

\begin{widetext}
\begin{equation}
W=\int_{V_P}^{V_e} PdV=  jm_e c^2 \int_{V_P}^{V_e} V^{-1} dV \approx j m_e c^2 \alpha^{-1} = j\hbar c /r_e.
\label{eq:W}
\end{equation}
\end{widetext}

In the case of the isothermal expansion of an ideal gas consisting of a large ensemble of non-interacting atoms or molecules, the energy required for the work done is supplied by a heat reservoir with a well-defined temperature.  For the expansion of a single vacuum fluctuation such a thermal reservoir is not available but, in quantum mechanics, an energy $W$ can be borrowed for a time interval $\tau$ limited by the uncertainty principle:  
\begin{equation}
\tau \approx \hbar / 2W =r_e/ 2jc.
\end{equation}

By setting $j=1/2$, $\tau=r_e/c$ becomes precisely the time required for a relativistic expansion from volume $V_P$ to $V_e$.  Using this value of $j$ in relation (\ref{eq:W}), the work-energy, $W$, becomes $\hbar c/2r_e=m_e c^2/2\alpha$ and the equation of state (\ref{eq:PV}) becomes  $PV=m_e c^2/2$.  (It is interesting to note in passing that the energy $m_e c^2/2$  also appears in electromagnetism as a the result of integrating the energy density of the electric field due to a point-like quantum of charge, $e$, over the infinite volume outside the classical electron sphere of radius $r_e$ and volume, $V_e$.)

The following phenomenological interpretation of this ideal gas model and of relations (\ref{eq:1}) and (\ref{eq:VeVp}) is proposed: a vacuum fluctuation formed at the Planck scale, $L_P$,  undergoes a quantum transition, expanding to a volume $V_e$ containing a virtual quantum state with energy, $W=\hbar c/2r_e$. In effect, the uncertainty principle provides the work-energy required for this transition. It also determines the lifetime, $\tau=r_e/c$, of the virtual state, after which it makes a transition back to the Planck-scale, so that energy is eventually conserved. 

This model is now developed to provide an insight into the measured value of the very small, finite and positive cosmological constant, $\Lambda$, by evaluating the negative gravitational self-energy density, $\rho_G$, of the expanded quantum state.  It is then found that $\rho_G$ has the same magnitude as the positive ``dark energy'' density, $\rho_\Lambda=c^2 \Lambda /8\pi G$, within 10\% accuracy, as follows.  The gravitational self-energy, $U_G$, of a sphere of radius $R$ and uniform mass density is given by $U_G=-3GM^2/5R$  where $M$ is the total mass. Its gravitational self-energy density is $\rho_G=- 9GM^2/20\pi R^4$.  For the virtual state corresponding to the expanded vacuum fluctuation, we set $R = r_e$, $M=W/c^2=\hbar /2cr_e=m_e/2\alpha$ so that $U_G(R=r_e)=3Gm_e^2 / 20 \alpha^2 r_e$. Assuming, for simplicity, that the mass $M$ is distributed uniformly within a sphere of radius $r_e$, the following expression for $\rho_G$ is then obtained: 
\begin{equation}
\frac{\rho_G}{c^2} =-\frac{9Gc^2 m_e^6}{80\pi \hbar^4 \alpha^6 }=-0.66 \times 10^{-26} \textrm{ kg m}^{-3}.
\label{eq:rG}
\end{equation}

According to the astronomical data in ref. \cite{20Patrignani}, the dark energy density corresponding to $\Lambda$ is given by 

\begin{equation}
\frac{\rho_\Lambda}{c^2}=0.60\times 10^{-26} \textrm{ kg m}^{-3}.
\label{eq:rL}
\end{equation}
(Since the present uncertainty in this value of $\rho_\Lambda$ is $\sim 4\%$, the values of the densities in relations (\ref{eq:rG}) and (\ref{eq:rL}) are given to 2 significant figures only).

Given the near coincidence of the magnitudes of these two very mass small densities, it is proposed that they are related physically so that
\begin{equation}
\rho_G+\rho_\Lambda \approx 0.
\label{eq:eqzero}
\end{equation}

By combining relations (\ref{eq:rG}) and (\ref{eq:eqzero}) the following expressions inter-relating the cosmological constant with five other fundamental constants is obtained: 
\begin{equation}
 \Lambda (\rho_\Lambda  \approx - \rho_G)\approx \frac{9G^2 c^2 m_e^6}{10\hbar^4 \alpha^6 }=1.11 \times 10^{-35} \textrm{ s}^{-2}.
 \label{eq:Lamb}
\end{equation}

This relation can also be written as

\begin{equation}
 \Lambda (\rho_\Lambda  \approx - \rho_G)\approx \frac{9c^2 l_p^4}{10 r_e^6}
 \label{eq:Lambred}
\end{equation}
where $l_p=\sqrt{\hbar G/c^3}$ is the ``reduced'' Planck length.  The numerical prefactor $9/10$ depends on the assumption that the energy $U_G$ is uniformly distributed within a sphere of radius $r_e$.  It could be ``fine-tuned'' to unity by adjusting slightly the mass-energy distribution.  

Relations (\ref{eq:rG}) to (\ref{eq:eqzero}) suggest that the negative gravitational potential energy $U_G(R=r_e)$ of the expanded vacuum fluctuation balances precisely the small and positive ``dark energy'' contained within its volume $V_e$.  Also, at $r_e=R$, the small and constant expansive pressure due to the positive value of $\Lambda$ is balanced by the compressive pressure of the gravitational self-energy, $U_G$.   Then, following the decay of the virtual state in a time $\tau=r_e/c$ required by the uncertainty principle, the volume $V_e$ generated by the expansion of the vacuum fluctuation contains only the dark energy given by the finite value of $\Lambda$ in relation (\ref{eq:Lamb}).  

\section{Discussion and conclusions}

A number of other research articles \cite{17Beck,21Wesson,22Wei,23Bohmer,24Nottale} have described methods to derive expressions for $\Lambda$ which are similar in form to relation (\ref{eq:Lamb}) but which are based on quite different physical arguments; see reference \cite{22Wei} for a recent review.  In particular, a statistical analysis by Beck \cite{17Beck} based on Kinchin axioms gives an expression very similar to (\ref{eq:Lamb}), but with unity replacing the numerical prefactor of 9/10. These evaluations of $\Lambda$ involve only $\hbar$ and the constants of the long-range gravitational and electromagnetic fields.  They suggest a fundamental inter-relation between gravity, quantum electrodynamics and the cosmological constant which can also be expressed in terms of the gravitational self-energy, $U_G$ of the expanded vacuum fluctuation:
\begin{equation}
\hbar \Lambda^{\frac{1}{2}}\sim U_G(R=r_e).
\end{equation}
This relation is suggestive of the uncertainty principle: since $\Lambda^{-1/2}$ is approximately equal to the present age of the universe, $T_u$, it seems to imply that the detection of a physical effect of such a small self-energy $U_G(R=r_e)$ requires a very large measurement time, $\sim T_u$.  Note also that the possibility of gravitational self-energy playing a role in quantum state reduction has been proposed by Penrose \cite{27Penrose,28Penrose,29Penrose}.    

The phenomenological model described here may be relevant to considerations of the anthropic principle \cite{25Carter,26Carter,3Carr,6Barrow}, and the concept of a multiverse \cite{27Tegmark,28Carr,29Ellis}  Since expressions (\ref{eq:1}) and (\ref{eq:Lamb}) inter-relate $\alpha$, $G$ and $\Lambda$, the apparent fine tuning of just one of these constants would seem to ensure the fine tuning of the other two.  Thus, the numerical value of  $\alpha \approx 1/137$ in the low energy limit, which underpins atomic and condensed matter physics, chemistry and biochemistry, implies a value of $G$, given by relation (\ref{eq:1}), that is large enough for galaxies, stars and planets to form, yet small enough for stellar life-times sufficiently long to enable the biological evolution of observers.  Similarly, the value of $\Lambda$ given by relation (\ref{eq:Lamb}) is large enough to be measured by astronomical equipment during the present epoch, yet small enough to provide sufficient time for the formation of galaxies. 

\section{Acknowledgements}
The author thanks  M.T. Greenaway, R.J.A. Hill, T.M. Fromhold, E.J. Copeland, P. Saffin and A. Armour for helpful discussions.

\end{document}